\documentclass[twocolumn,showpacs,preprintnumbers,amsmath,amssymb,aps]{revtex4}
\usepackage{graphicx}
\usepackage{dcolumn}
\usepackage{bm}

\begin{document}

\bibliographystyle{prsty}

\title{Subnatural linewidth using electromagnetically induced
transparency in Doppler-broadened vapor}
\author{S. M. Iftiquar, G. R. Karve, and Vasant Natarajan}
\email{vasant@physics.iisc.ernet.in}
\homepage{www.physics.iisc.ernet.in/~vasant}
\affiliation{Department of Physics, Indian Institute of
Science, Bangalore 560\,012, INDIA}

\begin{abstract}
We obtain subnatural linewidth (i.e.\ $<\Gamma$) for probe
absorption in room-temperature Rb vapor using
electromagnetically induced transparency (EIT) in a
$\Lambda$ system. For stationary atoms, the EIT dip for a
resonant control laser is as wide as the control Rabi
frequency $\Omega_c$. But in thermal vapor, the moving
atoms fill the transparency band so that the final EIT dip
remains subnatural even when $\Omega_c > \Gamma$. We
observe linewidths as small as $\Gamma/7$ in the $D_2$ line
of Rb.
\end{abstract}



\maketitle

Laser spectroscopy in a room-temperature gas is often
limited by Doppler broadening due to the thermal velocity
of gas particles. While techniques such as
saturated-absorption spectroscopy can be used to eliminate
the first-order Doppler effect and get linewidths close to
the natural linewidth, the natural linewidth itself appears
as a fundamental limit to the resolution that can be
achieved in precision spectroscopy. In addition, when
lasers are locked to atomic transitions (for use as
frequency standards), the natural linewidth determines the
tightness of the lock. It is therefore desirable to develop
techniques for getting below the natural linewidth.

In this work, we propose and demonstrate a technique to
obtain subnatural linewidth in a Doppler broadened medium.
The technique has been adapted from recent developments in
the use of control lasers in three level systems as a means
of modifying the absorption properties of a probe beam
\cite{NSO90}, in what is generally called coherent-control
spectroscopy. More specifically, we use the phenomenon of
electromagnetically induced transparency (EIT) in a
$\Lambda$-type system, in which an initially absorbing
medium is rendered transparent to a weak probe when a
strong control laser is applied to a second transition
\cite{BIH91}. It is well known that the EIT dip on
resonance for stationary atoms can be subnatural if the
Rabi frequency of the control laser is sufficiently small
\cite{LIX95}. However, in thermal vapor, the effect of the
large Doppler width was thought to have a detrimental
effect on observing any subnatural features. Indeed,
theoretical work in such Doppler broadened media predicted
that one can at best achieve sub-{\it Doppler} resolution
by detuning the control laser \cite{VAR96}. However, in
earlier work \cite{RWN03}, we have shown that we can obtain
sub-{\it natural} linewidth either by detuning the control
by an amount that is larger than the Doppler width, or by
using a slightly-detuned control along with a
counterpropagating pump beam that allows the probe to
address only zero-velocity atoms. Here, we show that one
can observe subnatural linewidth for the EIT dip even when
the control is on resonance.

There has been a previous report of subnatural linewidth
for the EIT dip using a $\Lambda$ system in the $D_1$ line
of Rb \cite{LIX95}. However, the dip was subnatural only
when the control Rabi frequency was less than the natural
linewidth, $\Gamma$, and reached a smallest value of
$\Gamma/4$ partly limited by the 3-MHz linewidth of the
laser. By contrast, we will see that our linewidth is about
$\Gamma/6$ even when the Rabi frequency is 5 times
$\Gamma$.

We must also contrast this with the related phenomenon of
coherent-population trapping (CPT) in $\Lambda$ systems
\cite{WYN99}, where the linewidth can be extremely small
because it is limited only by the decoherence rate between
the two ground levels. In contrast to EIT experiments, CPT
experiments require (i) the use of phase coherent control
and probe beams, (ii) roughly equal powers in the two
beams, (iii) detuning of the two beams (equally) from
resonance to decrease the decohering effect of the excited
state, and (iv) buffer gas filled vapor cells to increase
the ground coherence time. Under these conditions, the
control and probe beams pump the atoms into a dark
nonabsorbing state and probe transmission shows a narrow
peak, with a linewidth of 50 Hz being observed in the case
of Cs \cite{BNW97}. By contrast, EIT occurs because of the
AC Stark shift of the excited state by the strong control
laser. On resonance, this creates two dressed states
\cite{COR77} that are shifted equally from the unperturbed
level, and the probe absorption again shows a minimum at
line center. Since CPT is a ground-state coherence
phenomenon, it is effectively decoupled from the excited
state and is used for spectroscopy on the ground hyperfine
interval, which in the case of Cs is used in the SI
definition of the second. Indeed, the linewidth of 50 Hz is
not really subnatural because the natural linewidth of the
upper ground level is $<1$ Hz. On the other hand, we have
shown that the subnatural linewidth of EIT can be used for
high-resolution hyperfine spectroscopy on the {\it excited}
state \cite{RAN02}.

\begin{figure}
\resizebox{0.75\columnwidth}{!}{\includegraphics{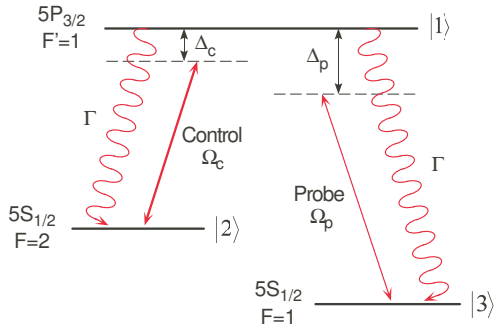}}
\caption{(Color online) Three-level $\Lambda$ system in the
$D_2$ line of $^{87}$Rb.} \label{levels}
\end{figure}

We now turn to the theoretical analysis of the three level
$\Lambda$ system shown in Fig.\ \ref{levels}. For
specificity, we show the relevant hyperfine levels in the
$D_2$ line of $^{87}$Rb. The strong control laser drives
the $|1 \rangle \leftrightarrow |2 \rangle$ transition with
Rabi frequency $\Omega_c$ and detuning $\Delta_c$, while
the weak probe is scanned across the $|1 \rangle
\leftrightarrow |3 \rangle$ transition. The spontaneous
decay rate from the upper state to either of the ground
states is $\Gamma$, which is $2\pi \times 6$ MHz in Rb. The
absorption of the weak probe is well known from a density
matrix analysis of the system \cite{LIX95R,VAR96}. It is
given by $-{\rm Im}(\rho_{13}\Gamma/\Omega_p)$, where
$\rho_{13}$ is the induced polarization on the $|1 \rangle
\leftrightarrow |3 \rangle$ transition. In the weak probe
limit, all the atoms will get optically pumped into the $|3
\rangle$ state, so that $\rho_{33} \approx 1$ and
$\rho_{22} \approx 0$. From the steady state solution of
the density matrix equations, we get (to first order in
$\Omega_p$),
\begin{equation}
\rho_{13}=-\frac{i\Omega_p/2} {\Gamma - i \Delta_p + i
\frac{\displaystyle |\Omega_c/2|^2}{\displaystyle
\Delta_p-\Delta_c}}.
\end{equation}
The pole structure of this equation shows that there will
be a zero in the absorption when $\Delta_p = \Delta_c$, and
this minimum will occur exactly on resonance if the control
is on resonance. This is the phenomenon of EIT.

The above analysis is correct for a stationary atom. For an
atom moving along the direction of the beams with a
velocity $v$, the detuning of the two beams will change by
$\pm kv$, where $k$ ($=2\pi/\lambda$) is the photon
wavevector and the sign depends on whether the atom is
moving away from or towards the beams. Thus, to obtain the
complete probe absorption in a gas of moving atoms, the
above expression has to be corrected for the velocity of
the atom and then averaged over the one-dimensional
Maxwell-Boltzmann distribution of velocities.

\begin{figure}
\resizebox{0.90\columnwidth}{!}{\includegraphics{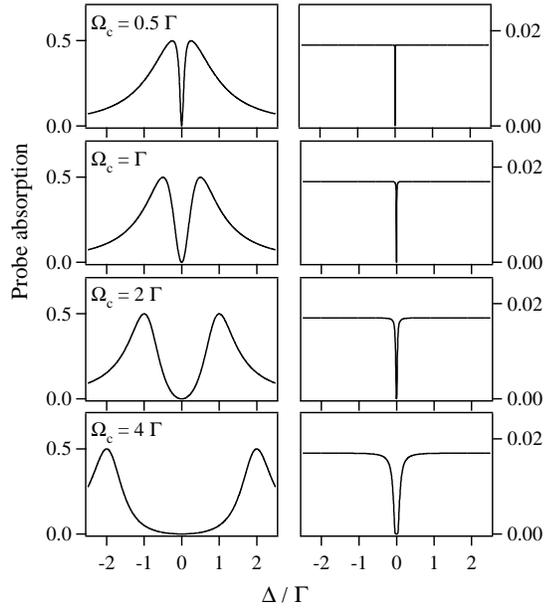}}
\caption{Calculated probe absorption for four values of
$\Omega_c$. The curves on the left are for zero-velocity
atoms, while the curves on the right are after thermal
averaging in room-temperature vapor. Note the decreased
scale on the right.} \label{doppler}
\end{figure}

The results of such a calculation for room-temperature Rb
atoms with $\Delta_c = 0$ and different values of
$\Omega_c$ are shown in Fig.\ \ref{doppler}. First let us
look at the curves for zero-velocity atoms shown on the
left. As is well known, the absorption splits into an
Autler-Townes doublet and shows a classic EIT dip in the
center. The doublet peaks are the two symmetric dressed
states created by the control laser \cite{COR77}, and their
separation is exactly equal to the value of $\Omega_c$.
Thus, the linewidth of the EIT dip (defined as the full
width at half maximum) depends on the value of $\Omega_c$.
It is subnatural (i.e.\ $<\Gamma$) only when $\Omega_c$ is
less than $\Gamma$, and arises because the absorption on
resonance goes to zero due to quantum interference between
the dressed states \cite{LIX95}. But the linewidth
increases quickly when $\Omega_c$ is increased, so that the
width is $3\Gamma$ when $\Omega_c=4\Gamma$.

Now consider the probe response after thermal averaging
shown on the right. The scale of absorption has decreased
by a factor of 30 as the absorption spreads over the
different velocity groups, {\it but the linewidth of the
EIT dip remains extremely small}. Indeed it only increases
to $\Gamma/6$ when $\Omega_c=4\Gamma$, and remains
subnatural for much higher values of $\Omega_c$.

\begin{figure}
\resizebox{0.85\columnwidth}{!}{\includegraphics{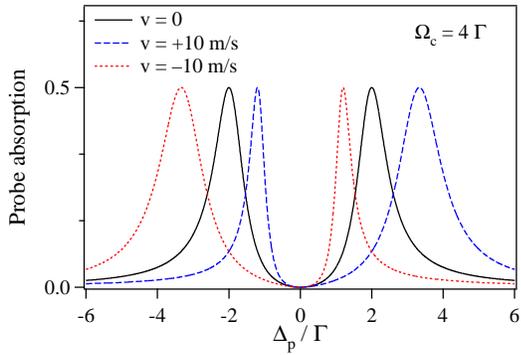}}
\caption{(Color online) Effect of velocity on probe
absorption. The curves are for zero velocity and for atoms
moving with 10 m/s to the right and left.} \label{vel}
\end{figure}

The prediction that the linewidth of the EIT dip after
thermal averaging becomes smaller is both surprising and
counter-intuitive. This can be understood better by
considering the effect of velocity on the EIT lineshape as
shown in Fig.\ \ref{vel} for $\Omega_c=4\Gamma$. The solid
curve is for stationary atoms, while the dashed (dotted)
curve is for atoms moving with $v=+10$ ($-10$) m/s. The
Autler-Townes doublet for the moving atoms is shifted to
the right or left so that they fill in the transparency
region for stationary atoms. The overall transparency
window shrinks and the effective EIT linewidth decreases.
Note that Fig.\ \ref{doppler} shows that this linewidth
reduction is accompanied by a change in the EIT lineshape
as well.

Such a surprising reduction in linewidth after thermal
averaging also happens for EIT in a ladder-type system.
This has been predicted and observed by us in earlier work
with room-temperature Rb atoms \cite{KPW05}. As in this
case, thermal averaging results in two additional features
(i) the scale of the transparency is reduced, and (ii) the
lineshape is modified from that for zero velocity atoms.
Our prediction of a modified lineshape showing enhanced
absorption near resonance has been recently observed for
EIT with Rydberg atoms \cite{MJA07}.

\begin{figure}
\resizebox{0.85\columnwidth}{!}{\includegraphics{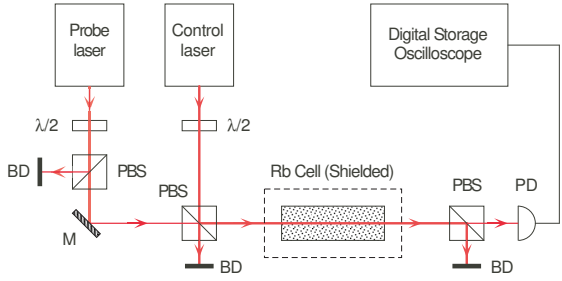}}
\caption{(Color online) Schematic of the experiment. Figure
key -- $\lambda/2$: halfwave plate, PBS: polarizing beam
splitter, BD: beam dump, M: mirror, PD: photodiode.}
\label{schematic}
\end{figure}

We now turn to the experimental demonstration of these
results. The experimental schematic is shown in Fig.\
\ref{schematic}. The probe and control beams are derived
from {\it independent} home-built frequency-stabilized
diode laser systems tuned to the 780 nm $D_2$ line of Rb
\cite{BRW01}. The linewidth of the lasers after
stabilization is about 1 MHz. The two beams are about 2 mm
in diameter each. The beams copropagate through a
room-temperature vapor cell with orthogonal polarizations.
The cell has a magnetic shield around it so that the
residual field (measured with a three-axis fluxgate
magnetometer) is $\sim$5 mG.

\begin{figure}
\resizebox{0.90\columnwidth}{!}{\includegraphics{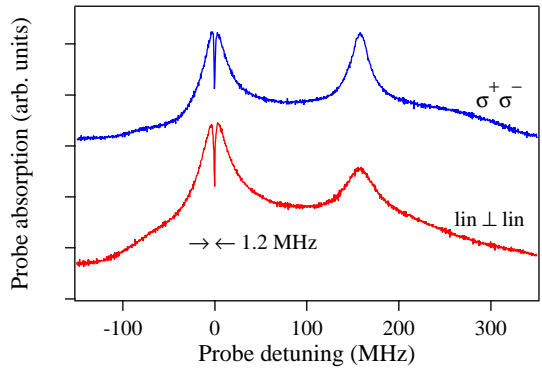}}
\caption{(Color online) Probe absorption showing a narrow
EIT dip for orthogonal circular ($\sigma^+\sigma^-$) and
linear (lin $\perp$ lin) polarizations. Probe detuning is
measured from the $F'=1$ level.} \label{eit1}
\end{figure}

In Fig.\ \ref{eit1}, we show a typical probe absorption
spectrum taken with a control power of 0.21 mW,
corresponding to a Rabi frequency of about 9 MHz ($1.5
\Gamma$), and a probe power of 0.07 mW. The most striking
feature of the curve is the 1.2 MHz ($\Gamma/5$) wide EIT
dip at line center, as predicted by our analysis above. The
signal-to-noise ratio of the dip is more than 20, and it
appears exactly at line center. Note that the control laser
is locked to the $F=2 \rightarrow F'=1$ transition using
standard saturated-absorption spectroscopy, where the
observed transition linewidths are about 15 MHz. Therefore,
one expects residual frequency jitter of the lock point on
the order of a few hundred kHz. Furthermore, the linewidth
of the probe laser is of the same order as the calculated
EIT linewidth. Despite these broadening effects, the
observed dip is only 1.2 MHz wide. The narrow line is quite
robust and appears whether the two beams have orthogonal
linear or circular polarizations, as seen from the figure.

Another interesting feature of the spectrum is that the
narrow EIT dip appears in the middle of a 40 MHz wide peak.
A similar peak also appears at a frequency 157 MHz higher,
which is precisely the hyperfine interval between the
$F'=1$ and $F'=2$ levels of the excited state. We thus
conclude that this peak is due to additional
velocity-dependent optical pumping by the control laser.
Note that the earlier density matrix analysis assumes
complete optical pumping by the strong control laser
\cite{LIX95R}. However such population transfer will be
true mainly for zero-velocity atoms for which the control
laser is on resonance, but will not be very effective for
moving atoms for which the control is detuned, particularly
when we consider that the probe has finite power and is no
longer in the weak probe limit. Thus the optical pumping
will be velocity dependent, and the population as a
function of velocity will deviate from the
Maxwell-Boltzmann distribution with a peak near zero
velocity.

\begin{figure}
\resizebox{0.85\columnwidth}{!}{\includegraphics{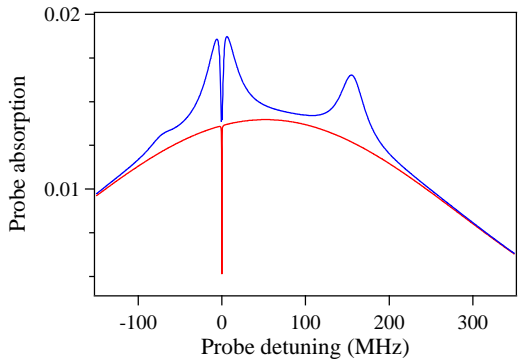}}
\caption{(Color online) Calculated probe absorption
spectrum in room temperature vapor. The lower curve is for
a Maxwell-Boltzmann velocity distribution, while the upper
curve is obtained after taking into account additional
optical pumping by the control laser for zero velocity
atoms.} \label{optpump}
\end{figure}

In Fig.\ \ref{optpump}, we compare the calculations of
probe absorption in room-temperature vapor with and without
such optical pumping. The velocity-dependent optical
pumping is incorporated phenomenologically by assuming a
population transfer that is proportional to the scattering
rate for control photons. The peak near zero in the
velocity distribution will show up as two additional peaks
in the probe absorption spectrum, one when these atoms come
into resonance with the $F=1 \rightarrow F'=2$ transition
(at $\Delta_p=+157$ MHz) and other for the $F=1 \rightarrow
F'=0$ transition (at $\Delta_p=-72$ MHz). A close
examination of the observed spectrum in Fig.\ \ref{eit1}
does show a small third peak at $\Delta_p=-72$ MHz. As
expected, the size of the optical pumping effect and the
relative height of the three broad peaks are different for
the two polarizations, but the lineshape is close to the
calculated one.

Further consideration of velocity-dependent optical pumping
shows that there will be a second velocity class that will
get optically pumped into the $F=1$ level. This happens for
atoms moving towards the control beam with a velocity near
122 m/s (corresponding to a Doppler shift of 157 MHz), for
which the control appears resonant with the $F=2
\rightarrow F'=2$ transition. This will cause two
additional peaks in the probe absorption, at
$\Delta_p=-157$ and $-229$ MHz. Since such optical pumping
is a competition between probe power and control detuning,
we see these additional peaks in our measured spectra only
when the probe power is reduced considerably.

\begin{figure}
\resizebox{0.95\columnwidth}{!}{\includegraphics{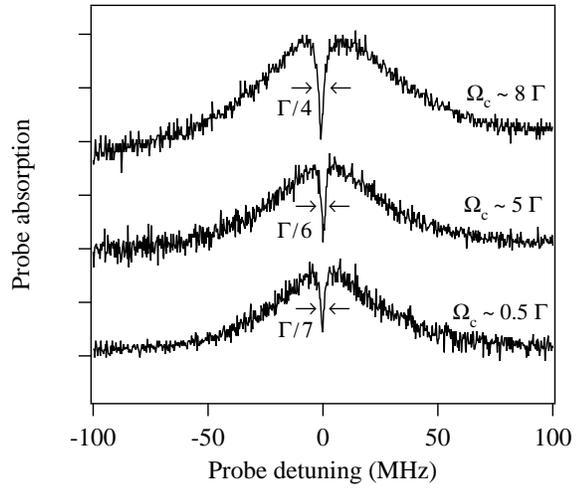}}
\caption{Dependence of EIT linewidth on control power. The
control Rabi frequency for each curve is indicated.}
\label{power}
\end{figure}

We finally consider the effect of control power on the EIT
linewidth. From our theoretical analysis, the EIT dip
should remain subnatural for quite high values of Rabi
frequency. This is confirmed from the three spectra shown
in Fig.\ \ref{power} taken with control powers of 0.02, 2,
and 6.5 mW, respectively. As the power is increased, the
overall size of the broad optical pumping peak increases.
But the linewidth of the EIT dip only increases from 0.88
MHz ($\Gamma/7$) to 1.5 MHz ($\Gamma/4$). This clearly
demonstrates the validity of our prediction that the
linewidth in thermal vapor is well below the control Rabi
frequency, or equivalently the separation of the dressed
states.

In conclusion, we have predicted and demonstrated
subnatural width for the EIT dip in thermal Rb vapor. While
the transparency band for stationary atoms is of the order
of the Rabi frequency of the control laser, in thermal
vapor the moving atoms fill up this band in such a manner
that the residual EIT dip remains extremely narrow even for
large values of Rabi frequency. The observed lineshape is
described well by a density-matrix treatment of the
three-level system with thermal averaging. The most
important advantage of the narrow feature over our previous
work with detuned control lasers \cite{RWN03} is that the
dip appears exactly at line center. Thus the feature can be
used for high resolution spectroscopy and tight laser
locking. The narrow feature is robust in terms of laser
polarization and detuning. This could be important for
applications of EIT such as nonlinear optics, gain without
inversion, slowing of light, and quantum information
processing.

\begin{acknowledgments}
This work was supported by the Department of Science and
Technology, India. One of us (V.N.) would like to
acknowledge useful discussions with E. Arimondo, G. M.
Saxena, and K. Pandey. He is the recipient of a Homi Bhabha
fellowship and the others (S.M.I. and G.R.K.) acknowledge
financial support from the Council of Scientific and
Industrial Research, India.
\end{acknowledgments}


\end{document}